\documentclass{article}
\usepackage{amssymb,amsmath}
\usepackage{graphicx}

\input{tcilatex}
\begin{document}

\title{Planck mass and Dilaton field as a function of the noncommutative
parameter}
\author{S. Zaim and Z. Aouachria \\
Department of Physics, University of Batna 05000, Algeria}
\date{}
\maketitle

\begin{abstract}
A deformed Bianchi type I metric in noncommutative gauge gravity is
obtained. The gauge potential (tetrad fields) and scalar curvature are
determined up to the second order in the noncommutativity parameters. The
noncommutativity correction to the Einstein-Hilbert action is deduced. We
obtain the Planck mass, on noncommutative space-time as a function of the
noncommutative parameter $\theta $, which implies that noncommutativity has
modified the structure and topology of the space-time.

\textbf{\emph{Keywords}}: \emph{Noncommutative field theory ,Gauge field
theory, Quantum gravity, Quantum fields in curved spacetime.}

PACS: 11.10.Nx ,11.15.-q , 04.60.-m , 04.62.+v
\end{abstract}

\section{Introduction}

Matter and field are basic concepts of classical field theories. They play a
fundamental role in the general relativity theory $[1]$, where the Einstein
tensor $G_{\mu }^{\nu }$ is expressed in terms of the geometry of
space-time, and the matter is represented by its momentum energy density
tensor $T_{\mu }^{\nu }$. These two intrinsic concepts are connected by the
Einstein field equation 
\begin{equation}
G_{\mu }^{\nu }=-8\pi T_{\mu }^{\nu }.
\end{equation}

According to eq. $(1)$, a given distribution of matter (sources) determines
the geometric properties of space-time. One can regard this as the creation
of space-time geometry by matter. Now, one can read eq. $(1)$ in the
opposite direction, and expect the creation of matter by geometry, which is
an interesting mechanism for creating particles.

The standard concept of space time as a geometric manifold is based on the
notion of points being locally labelled by a finite number of real
coordinates. However, it is generally believed that this picture of space
time as a manifold should break down at very short distances of the order of
the Planck length. This implies that the mathematical concepts of high
energy physics have to be changed or more precisely our classical geometry
concepts may not be well suited for the description of physical phenomenon
at short distances.

Noncommutativity is a mathematical concept expressing uncertainty in quantum
mechanics, where it applies to any pair of conjugate variables such as
position and momentum. It would be interesting to look for a mechanism of
creating particles in the noncommutative geometry. There are several
motivations to speculate that space-time becomes non-commutative at very
short distances when quantum gravity becomes relevant. Moreover, in string
theories, the non-commutative gauge theory appears as a certain limit in the
presence of a background field $[2-3]$. In this context, a gauge field
theory with star products and Seiberg-Witten maps is used .

We shall work with the noncommutative canonical space-time, which can be
realized by the generalisation of the commutation relation for the canonical
variables (coordinate-momentum operators) to non-trivial ones of coordinate
operators. It was shown that combing Heisenberg uncertainty principle with
Einstein theory of classical gravity leads to the conclusion that ordinary
space-time loses any operational meaning at short distances, that is a
space-time coordinate with a great accuracy $\epsilon $ causes an
uncertainty in momentum or energy of the order of $\epsilon ^{-1}$ which is
transmitted to the system and concentrated at some time in the localization
point . Exploring the limitations of localisation measurements which are due
to the possible black hole creation by concentration of energy, one arrives
at uncertainty relations among space-time coordinates which can be traced
back to the noncommutative commutation relations: 
\begin{equation}
\left[ \hat{x}^{\mu } , \hat{x}^{\nu }\right] _{\ast }=i\theta ^{\mu \nu },
\end{equation}
where $\hat{x}$ are the coordinate operators and $\theta ^{\mu \nu }$ is the
noncommutativity parameter and is of dimension of length-squared, and
corresponds to the smallest patch of area in physical space one may
``observe'', similar to the role $\hbar $ plays in \ $\left[ \hat{x}_{i}, 
\hat{p}_{j}\right] =i\hbar \delta _{ij}$, defining the corresponding
smallest patch of phase space in quantum mechanics.

It is a challenge to formulate a theory of gravitation on a non-commutative
manifold. The main problem lies in the fact that it is difficult to
implement symmetries such as general coordinate covariance and local Lorentz
invariance and to define derivatives which are torsion-free and satisfy the
metricity condition. In a flat space-time, to get noncommutative local gauge
theories with Lorentz violation symmetry, a formulation within the
enveloping algebra approach has been proposed $\left[ 2\right] -\left[ 4 %
\right] $. Following a similar path, a gauge formulation of gravity is
proposed $\left[ 3-5-6\right] $. It is a theory of general relativity on
curved spacetime with canonical preservation of noncommutative spacetime
commutation relations and based partially on implementing symmetries on flat
noncommutative spacetime. The non-locality in time-like noncommutativity $%
\theta ^{0i}\left( i=1,2,3\right) $ leads to the unitarity violation $\left[
7\right] .$ To overcome this problem we obtain the generalised local Lorentz
and general coordinates infinitesimal transformations of the noncommutative
space-time, which are exact symmetries of the canonical noncommutative
spacetime commutation relations without any further constraints like
unimodularity $\left[ 8\right] $.

In this paper, we present a deformed Bianchi type I metric in
non-commutative gauge theory of gravitation. Then using the Seiberg-Witten
map we calculate the non-commutative corrections for the gauge potentials
(tetrad fields) up to the second order of the expansion in $\theta $. Using
the results we determine the deformed Bianchi type I metric. The correction
is obtained up to the second order of the noncommutativity parameter, and we
compute the Ricci scalar curvature in a noncommutative cosmological
anisotropic Bianchi I universe. The noncommutative correction to the
Einstein-Hilbert action is deduced.

\section{The gauge theory}

The anisotropic Bianchi type I space-time metric is expressed by the
line-element: 
\begin{equation}
ds^{2}=-dt^{2}+t^{2}\left( dx^{2}+dy^{2}\right) +dz^{2}.
\end{equation}
The corresponding metric $g_{\mu \nu }$ has the following non-zero
components: 
\begin{equation}
g_{00}=-1,g_{11}=t^{2}=g_{22},g_{33}=1.
\end{equation}
The tetrads $e_{\mu }^{a}(x),a=0,1,2,3,$ and the spin connections $\omega
_{\mu }^{ab}(x)=-\omega _{\mu }^{ba}(x),$ $%
[ab]=[01],[02],[03],[12],[13],[23] $, are written as follows: 
\begin{equation}
\omega _{\mu }^{ab}=e_{\rho }^{a}e^{\nu b}\Gamma _{\mu \nu }^{\rho }-e^{\nu
b}\partial _{\mu }e_{\nu }^{a}.
\end{equation}
Here $e_{a}^{\nu }(x)$ denotes the inverse of $e_{{\mu}}^{a}(x)$ satisfying
the usual properties: 
\begin{equation}
e_{{\mu}}^{a}e_{b}^{\mu }=\delta _{b}^{a},e_{ {\mu}}^{a}e_{a}^{\nu }=\delta
_{\mu }^{\nu },
\end{equation}
and $\Gamma _{\mu \nu }^{\rho }$ is the affine connection, which is written
in function of the metric $g_{\mu \nu }$ as: 
\begin{equation}
\Gamma _{\mu \nu }^{\rho }=\frac{1}{2}g^{\rho \sigma }\left( \partial _{\mu
}g_{\sigma \nu }+\partial _{\nu }g_{\sigma \mu }-\partial _{\sigma }g_{\mu
\nu }\right) .
\end{equation}

Then, the corresponding components of the strength tensor can be written in
the standard form as the torsion tensor: 
\begin{equation}
R_{\mu \nu }^{a}=\partial _{\mu }e_{\nu }^{a}-\partial _{\nu }e_{\mu
}^{a}+\left( \omega _{\mu }^{ab}e_{\nu }^{d}+\omega _{\nu }^{ab}e_{\mu
}^{d}\right) \eta _{bd},
\end{equation}
with $\eta _{bd}$ the flat space metric, and the curvature tensor: 
\begin{equation}
R_{\mu \nu }^{ab}=\partial _{\mu }\omega _{\nu }^{ab}-\partial _{\nu }\omega
_{\mu }^{ab}+\left( \omega _{\mu }^{ac}\omega _{\nu }^{db}+\omega _{\nu
}^{ac}\omega _{\mu }^{db}\right) \eta _{cd}.
\end{equation}

The particular form of the line-element gravitational gauge field is given
by the following ansatz: 
\begin{equation}
e_{\mu }^{0}=\left( 1,0,0,0\right) ,e_{\mu }^{1}=\left( 0,t,0,0\right)
,e_{\mu }^{2}=\left( 0,0,t,0\right) ,e_{\mu }^{3}=\left( 0,0,0,1\right) ,
\end{equation}
and 
\begin{equation}
\Gamma _{11}^{0}=\Gamma _{22}^{0}=t,\Gamma _{10\nu }^{1}=\Gamma _{20}^{2}= 
\frac{1}{t},\omega _{1}^{01}=\omega _{2}^{02}=1.
\end{equation}
Here we only give the expressions of $R_{{\mu}\nu }^{ab}$ components, which
we need to use further in the derivation of the expressions of tetrads: 
\begin{equation}
R_{12}^{12}=R_{21}^{21}=R_{21}^{12}=-R_{12}^{21}=1.
\end{equation}
The scalar curvature is $R=R_{{\mu}\nu }^{ab}e_{a}^{\mu }e_{b}^{\nu
}=2/t^{2}.$

\section{Noncommutative gauge theory}

We assume that the noncommutative canonical structure of space time is
defined by eq. (1) where $\ast $ is the star product defined between the
functions $f$ and $g$ over this spacetime: 
\begin{equation}
\left( f\ast g\right) \left( x\right) =f\left( x\right) \exp \left( \frac{i}{%
2}\theta ^{\mu \nu }\overleftarrow{\partial }_{\mu }\overrightarrow{\partial 
}_{\nu }\right) g\left( x\right) .
\end{equation}

The gauge field for the noncommutative canonical space-time are denoted by $%
\hat{\omega}_{\mu }^{ab}\left( x,\theta \right) ,$ subject to the conditions$%
\left[ 9,10\right] :$ 
\begin{equation}
\hat{\omega}_{\mu }^{ab+}\left( x,\theta \right) =-\hat{\omega}_{\mu
}^{ba}\left( x,\theta \right) =\hat{\omega}_{\mu }^{ab}\left( x,-\theta
\right) =\hat{\omega}_{\mu }^{abr}\left( x,\theta \right) .
\end{equation}%
By expanding the gauge fields in powers of $\theta ^{2}$, 
\begin{equation}
\hat{\omega}_{\mu }^{ab}\left( x,\theta \right) =\omega _{\mu }^{ab}\left(
x\right) +\theta ^{\alpha \beta }\omega _{\mu \alpha \beta }^{ab}\left(
x\right) +\theta ^{\alpha \beta }\theta ^{\gamma \delta }\omega _{\mu \alpha
\beta \gamma \delta }^{ab}\left( x\right) +O\left( \theta ^{3}\right) ,
\end{equation}%
the above conditions then imply the following: 
\begin{equation}
\omega _{\mu }^{ab}\left( x\right) =-\omega _{\mu }^{ab}\left( x\right)
,\omega _{\mu \alpha \beta }^{ab}\left( x\right) =\omega _{\mu \alpha \beta
}^{ba}\left( x\right) .
\end{equation}%
Using the Seiberg-Witten map $\left[ 2\right] ,$ one obtains the following
noncommutative corrections up to the second order $\left[ 11\right] :$ 
\begin{eqnarray}
\omega _{\mu \alpha \beta }^{ab}\left( x\right)  &=&-\frac{i}{4}\left\{
\omega _{\alpha },\partial _{\beta }\omega _{\mu }+R_{\beta \mu }\right\}
^{ab}, \\
\omega _{\mu \alpha \beta \gamma \delta }^{ab}\left( x\right)  &=&\frac{1}{32%
}\Big(\left[ \omega _{\gamma },2\left\{ R_{\delta \alpha },R_{\mu \beta
}\right\} -\left\{ \omega _{\alpha },\left( D_{\beta }R_{\delta \mu
}+\partial _{\beta }R_{\delta \mu }\right) \right\} \right. -  \notag \\
&&\left. -\partial _{\delta }\left\{ \omega _{\alpha },\left( \partial
_{\beta }\omega _{\mu }+R_{\beta \mu }\right) \right\} \right] ^{ab}+\left[
\partial _{\alpha }\omega _{\gamma },\partial _{\beta }\left( \partial
_{\delta }\omega _{\mu }+R_{\delta \mu }\right) \right] ^{ab}-  \notag \\
&&\left. -\left\{ \left\{ \omega _{\alpha },\left( \partial _{\beta }\omega
_{\gamma }+R_{\beta \gamma }\right) \right\} ,\left( \partial _{\delta
}\omega _{\mu }+R_{\delta \mu }\right) \right\} ^{ab}\right) ,
\end{eqnarray}%
where 
\begin{equation}
\left\{ A,B\right\} ^{ab}=A^{ac}B_{c}^{b}+B^{ac}A_{c}^{b},\left[ A,B\right]
^{ab}=A^{ac}B_{c}^{b}-B^{ac}A_{c}^{b},
\end{equation}%
and 
\begin{equation}
R_{\mu \nu }^{ab}=\partial _{\mu }\omega _{\nu }^{ab}-\partial _{\nu }\omega
_{\mu }^{ab}+\omega _{\mu }^{ac}\omega _{\nu }^{cb}-\omega _{\nu
}^{ac}\omega _{\mu }^{cb},D_{\mu }R_{\alpha \beta }^{ab}=\partial _{\mu
}R_{\alpha \beta }^{ab}+\left( \omega _{\mu }^{ac}R_{\alpha \beta
}^{db}+\omega _{\mu }^{bc}R_{\alpha \beta }^{da}\right) \eta _{cd}\,.
\end{equation}%
The result for $\hat{e}_{\mu }^{a}$ up to the second order in $\theta $ is: 
\begin{eqnarray}
\hat{e}_{\mu }^{a} &=&e_{\mu }^{a}-\frac{i}{4}\theta ^{\alpha \beta }\left(
\omega _{\alpha }^{ac}\partial _{\beta }e_{\mu }^{c}+\left( \partial _{\beta
}\omega _{\mu }^{ac}+R_{\beta \mu }^{ac}\right) e_{\mu }^{c}\right)   \notag
\\
&&+\frac{1}{32}\theta ^{\alpha \beta }\theta ^{\gamma \delta }\left(
2\left\{ R_{\delta \alpha }R_{\mu \beta }\right\} ^{ac}e_{\gamma
}^{c}-\omega _{\gamma }^{ac}\left( D_{\beta }R_{\delta \mu }^{cd}+\partial
_{\beta }R_{\delta \mu }^{cd}\right) e_{\alpha }^{d}\right.   \notag \\
&&\left. -\left\{ \omega _{\alpha }\left( D_{\beta }R_{\delta \mu }+\partial
_{\beta }R_{\delta \mu }\right) \right\} ^{ad}e_{\gamma }^{d}-\partial
_{\delta }\left\{ \omega _{\alpha }\left( \partial _{\beta }\omega _{\mu
}+R_{\beta \mu }\right) \right\} ^{ac}e_{\gamma }^{c}\right.   \notag \\
&&\ \left. -\omega _{\gamma }^{ac}\partial _{\delta }\left\{ \omega _{\alpha
}\partial _{\beta }e_{\mu }^{d}+\left( \partial _{\beta }\omega _{\mu
}^{cd}+R_{\beta \mu }^{cd}\right) e_{\alpha }^{d}\right\} +\partial _{\alpha
}\omega _{\gamma }^{ac}\partial _{\beta }\partial _{\delta }e_{\mu
}^{c}\right.   \notag \\
&&\ \ \left. -\partial _{\beta }\left( \partial _{\delta }\omega _{\mu
}^{ac}+R_{\delta \mu }^{ac}\right) \partial _{\alpha }e_{\gamma
}^{c}-\left\{ \omega _{\alpha }\left( \partial _{\beta }\omega _{\gamma
}+R_{\beta \gamma }\right) \right\} ^{ac}\partial _{\delta }e_{\mu
}^{c}\right.   \notag \\
&&\ \left. -_{\beta }\left( \partial _{\delta }\omega _{\mu }^{ac}+R_{\delta
\mu }^{ac}\right) \left( \omega _{\alpha }^{cd}\partial _{\beta }e_{\gamma
}^{d}+\left( \partial _{\beta }\omega _{\gamma }^{cd}+R_{\beta \gamma
}^{cd}\right) e_{\alpha }^{d}\right) \right) +\mathcal{O}\left( \theta
^{3}\right) .
\end{eqnarray}

Then we can introduce a noncommutative metric by the formula: 
\begin{equation}
\hat{g}_{\mu \nu }=\frac{1}{2}\eta _{ab}\left( \hat{e}_{\mu }^{a}\ast \hat{e}%
_{\nu }^{b+}+\hat{e}_{\mu }^{b}\ast \hat{e}_{\mu }^{a+}\right) ,
\end{equation}%
where the $\hat{e}_{\mu }^{a+}$ is the complex conjugate of the
noncommutative tetrad fields given in $\left( 21\right) $ .

\subsection{Second order correction to Bianchi Type I}

We choose the coordinate system so that the noncommutative parameters $%
\theta ^{\alpha \beta }$ are given by $\left[ 12\right] $ : 
\begin{equation}
\theta ^{\alpha \beta }=\left( 
\begin{array}{cccc}
0 & 0 & 0 & \theta \\ 
0 & 0 & \theta & 0 \\ 
0 & -\theta & 0 & 0 \\ 
-\theta & 0 & 0 & 0%
\end{array}%
\right) .
\end{equation}

The nonzero components of the tetrad fields $\hat{e}_{\mu }^{a}$ are: 
\begin{eqnarray}
\hat{e}_{1}^{1} &=&t\left( 1-\frac{3}{32}\theta ^{2}\right) +\mathcal{O}%
\left( \theta ^{3}\right) , \\
\hat{e}_{2}^{1} &=&-i\frac{t}{4}\theta +\mathcal{O}\left( \theta ^{3}\right)
, \\
\hat{e}_{1}^{2} &=&i\frac{t}{4}\theta +\mathcal{O}\left( \theta ^{3}\right) ,
\\
\hat{e}_{2}^{2} &=&t\left( 1-\frac{3}{32}\theta ^{2}\right) +\mathcal{O}%
\left( \theta ^{3}\right) , \\
\hat{e}_{3}^{3} &=&1+\mathcal{O}\left( \theta ^{3}\right) , \\
\hat{e}_{0}^{0} &=&1+\mathcal{O}\left( \theta ^{3}\right) .
\end{eqnarray}%
Then, using the definition $\left( 22\right) ,$we obtain the following
non-zero components of the noncommutative metric $\hat{g}_{\mu \nu }$ up to
the second order of $\theta :$ 
\begin{eqnarray}
\hat{g}_{11} &=&t^{2}-\frac{3}{32}t^{2}\theta ^{2}+\mathcal{O}\left( \theta
^{3}\right) , \\
\hat{g}_{22} &=&t^{2}-\frac{3}{32}t^{2}\theta ^{2}+\mathcal{O}\left( \theta
^{3}\right) , \\
\hat{g}_{33} &=&1+\mathcal{O}\left( \theta ^{3}\right) , \\
\hat{g}_{00} &=&-1+\mathcal{O}\left( \theta ^{3}\right) ,
\end{eqnarray}%
such that for $\theta \rightarrow 0$ we obtain the ordinary ones. Then we
can calculate the noncommutative correction to the Ricci scalar curvature
given by: 
\begin{equation}
\hat{R}=\hat{e}_{a}^{\mu }\ast \hat{R}_{{\mu }\nu }^{ab}\ast \hat{e}%
_{b}^{\nu }=\frac{2}{t^{2}}\left( 1-\frac{3}{32}\theta ^{2}\right)
\thickapprox R\exp \left( -\frac{3}{32}\theta ^{2}\right) ,
\end{equation}%
where $R=2/t^{2}$ is the ordinary scalar curvature.

The Einstein-Hilbert action is: 
\begin{equation}
S_{g}=-\frac{1}{16\pi G}\int d^{4}xR\sqrt{-g}\exp \left( -\frac{3}{16}\theta
^{2}\right) .
\end{equation}%
When the effective coupling $\exp \left( \frac{3}{16}\theta ^{2}\right) $ is
small, then the Dilaton field is proportional to $\theta ^{2}$ $\left[ 13%
\right] $. In this case the action $\left( 35\right) $ can by written in
this form: 
\begin{equation}
S_{g}=-\frac{1}{16\pi G}\int d^{4}x\sqrt{-g}\left[ \hat{R}+W\left( \theta
^{2}\right) \right] ,
\end{equation}%
where the $W\left( \theta ^{2}\right) $ is the Dilaton potential. We note
through eq. (31) that the noncommutative geometry creates the Dilaton field
which is automatically within the gravity and we think that this new
coupling represents the source matter. The action $(35)$ as such also
undergoes a modification that is reflected by the coupling constant $\hat{k}%
=k\exp \left( \frac{3}{16}\theta ^{2}\right) $ ($k$ is the ordinary
coupling). This expression is the same as that when the action is written in
the space of five dimensions; then we can say that noncommutativity is
responsible of the fifth dimension of the gravity. With all this the
equation of motion derived from$(35)$ is: 
\begin{equation}
e^{-\frac{3}{16}\theta ^{2}}\left( R_{\mu \nu }-\frac{1}{2}Rg_{\mu \nu
}\right) =0.
\end{equation}

The equation $(37)$ takes the same form as in the or ordinary space- time
although it is multiplied by the factor $e^{-\frac{3}{16}\theta ^{2}}$which
has no effect on it. . However, this factor influences the coupling constant
k as we have seen previously. Thus the Planck mass is a function of the
parameter noncommuttive as follows:

\begin{equation}
\hat{M}_{p}=M_{p}\left( 1-\frac{3}{32}\theta ^{2}\right) \sim M_{p}\exp
\left( -\frac{3}{32}\theta ^{2}\right) ,
\end{equation}

where $M_{p}$ is the ordinary Planck mass. To get an idea about the
qualitative effect of noncommutativity, we display the ratio $M_{p}/\hat{M}%
_{p}$ as a function of the noncommutative parameter $\theta $ (see Fig.
below). Note that if the noncommutative parameter $\theta $ increases the
ratio $M_{p}/\hat{M}_{p}$ increases. These different values are essetially
characterised by different the gravitational constant values on
noncommutative space-time. This variation depends to the noncommutative
parameter $\theta .$Thus the Planck mass, in noncommutative space-time, is
composed of two parts: the ordinary one and the second part which is
negative and expressed in the second order of $\theta $. We understand that
this negative value represents the part of the matter which is kept secret
in the noncommutative coordinates; which can explain the dark matter.

\bigskip \FRAME{dtbpFUX}{3.5518in}{1.4062in}{0pt}{\Qcb{Plank mass ratio
versus theta}}{}{Plank mass ratio versus theta}{\special{language
"Scientific Word";type "MAPLEPLOT";width 3.5518in;height 1.4062in;depth
0pt;display "USEDEF";plot_snapshots TRUE;mustRecompute FALSE;lastEngine
"MuPAD";xmin "0";xmax "1";xviewmin "-0.00010000010002";xviewmax
"1.00010000010002";yviewmin "0.999990171475967";yviewmax "1.1";plottype
4;labeloverrides 3;x-label "theta";y-label "f(theta)";axesFont "Times New
Roman,12,0000000000,useDefault,normal";numpoints 100;plotstyle
"patch";axesstyle "normal";axestips FALSE;xis \TEXUX{v58130};var1name
\TEXUX{$\theta $};function \TEXUX{$e^{\frac{3}{32}\theta ^{2}}$};linecolor
"black";linestyle 1;pointstyle "point";linethickness 1;lineAttributes
"Solid";var1range "0,1";num-x-gridlines 100;curveColor
"[flat::RGB:0000000000]";curveStyle "Line";rangeset"X";VCamFile
'LDIMEN00.xvz';valid_file "T";tempfilename
'KVQ4ZY00.wmf';tempfile-properties "XPR";}}

\section{Conclusions}

Throughout this work we used Seiberg-Witten maps and the Moyal product up to
the second order of the noncommutativity parameter $\theta $, we choose the
Bianchi I universe and derive both the second order noncommutative
correction of the metric, and the deformed Einstein-Hilbert action. In
addition we obtained the Planck mass on noncommutative space-time as a
function of noncommutative parameter $\theta$, which implies that the
noncommutativity has modified the structure and topology of the space-time.
This induces the connection between the mass and parameter responsible of
noncommutative geometry which permits us to understand the dark matter and
dark energy especially if we take the noncommutative parameter dependent on
the coordinates.

\section{acknowledgement}

The authors thank Dr. Y. Delenda for useful discussions.

\end{document}